# Control of the surface plasmon dispersion and Purcell effect at the metamaterial-dielectric interface

Elizaveta I.Girshova [1,2], Konstantin A. Ivanov [1,2], Konstantin M. Morozov [1,2], Galia Pozina [3], and Mikhail A. Kaliteevski [1,2,4]

[1] ITMO University, 197101 St. Petersburg, Russia
[2] St. Petersburg Academic University, 194021 St. Petersburg, Russia
[3] Department of Physics, Chemistry and Biology (IFM), Linköping University, SE-58183 Linköping, Sweden
[4] Ioffe Institute, 194021 St. Petersburg, Russia

**Abstract**

The use of metamaterial as a way to mitigate the negative effects of absorption in metals on the Purcell effect in metal-dielectric structures is investigated. A layered metal-dielectric structure is considered as an anisotropic medium in the long-wavelength limit. The dispersion of the surface plasmon appearing at the boundary between such a structure and a different dielectric material, as well as the position of the peak in the local density of states are studied for various combinations of materials and filling factors of the periodic structure. The calculated frequency dependence of the Purcell factor demonstrates an increase in peak value compared to the conventional plasmonic structure.

Surface plasmon, a localized state of an electromagnetic field at the interface between a metal and a dielectric, was predicted more than sixty years ago.[1] Due to the formation of such states, metal-dielectric structures can facilitate a strong light-matter interaction and therefore attract significant research interest.[2-5] For example, field localization provides possibilities for the development of subwavelength optical devices,[6,7] while the increased amplitude of the field near the interface allows to utilize such systems in sensor devices[8,9]. Moreover, a number of spectacular effects caused by surface plasmons can be mentioned,[10-13] in particular, a surface-enhanced Raman scattering[14,15] and the Purcell effect[16], which is the enhancement of spontaneous emission probability in an inhomogeneous medium. The latter phenomenon is crucial for increasing the efficiency of light emission in optoelectronic and photonic devices.[17,18]

Previously, it was shown theoretically that the major enhancement of the spontaneous emission probability can be achieved in plasmonic structures due to a high local density of states (LDOS)[19]. However, this conclusion has been argued[20] in recently published paper[21], where it was demonstrated that for the frequency range, where the LDOS peak is occurring, this enhancement is dramatically reduced due to light absorption in metals. Nevertheless, the existence of features in the dispersion curve suggests that the effective utilization of the surface plasmons still can be achieved[22].

A possible way to obtain higher magnitudes of the enhancement of the spontaneous emission probability is to shift the LDOS peak to the low frequency range, where absorption is lower. Different structures have been proposed, where the properties of material depend on the geometry and are described by values such as effective plasma frequency[23,24]. These material systems include metallic wire structures[25] and various 3D-metamaterials[26-28], which are interesting but hard to model, fabricate and, most importantly, the control of dispersion in them is not obvious.

On the other hand, one-dimensional metal-dielectric structures are easy to manufacture, and at the same time, surface plasmon dispersion at the interface between such a metamaterial and a dielectric obeys quite simple rules, as was shown previously[29,30]. Thus, we investigate in this work surface plasmon states in one-dimensional metal-dielectric structures. In particular, we obtain the values of the LDOS peak frequency depending on the metamaterial filling factor and we calculate the possible values of the Purcell factor.

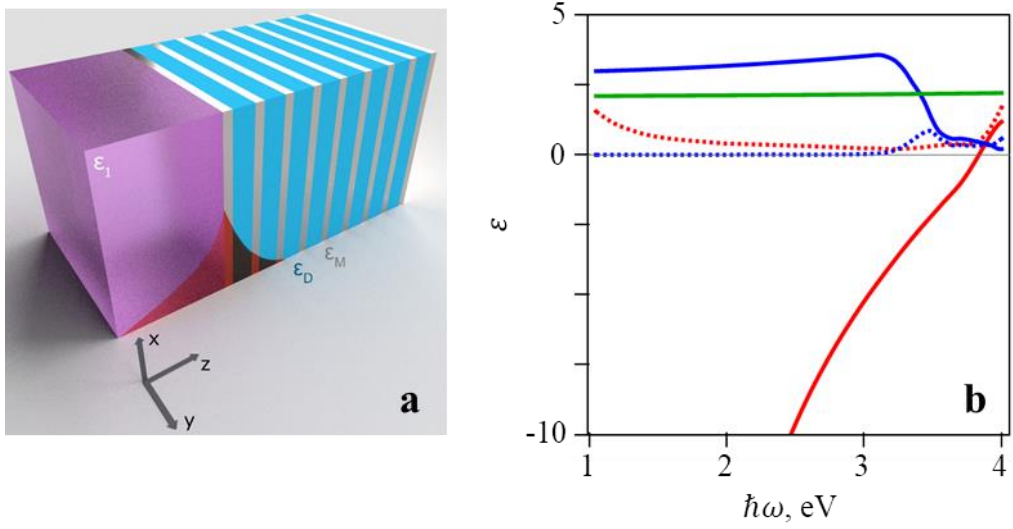

**Fig. 1:** (a) Geometry and composition of the structure. Silver and blue colors denote metallic and dielectric layers of metamaterial, respectively. The purple color denotes dielectric cladding. The red curve represents the electric field of the surface plasmon. (b) Dielectric constant dispersion in materials used in calculations (solid – real part, dashed – imaginary part): red – silver, blue – CBP, green – silica.

The proposed metamaterial is a one-dimensional structure consisting of alternating layers of metal and dielectric parallel to the $xy$ plane (see Figure 1). The dielectric functions are labeled by $\varepsilon_{Me}$ and $\varepsilon_D$, and the corresponding layer thicknesses by $a$ and $b$, respectively. We denote the fraction of metal in the metamaterial, i.e. the filling factor, as

$$\alpha = \frac{a}{a+b} \tag{1}$$

It is known that in the long-wavelength limit, which is satisfied for the low-frequency range, such a structure can be considered as a uniform anisotropic medium with a dielectric tensor:

$$\begin{pmatrix} \varepsilon_x & 0 & 0 \\ 0 & \varepsilon_x & 0 \\ 0 & 0 & \varepsilon_z \end{pmatrix} \tag{2}$$

Note that only two different non-zero components are present. They are defined as:

$$\varepsilon_x = \alpha \varepsilon_M + (1-\alpha)\varepsilon_D, \qquad \frac{1}{\varepsilon_z} = \frac{\alpha}{\varepsilon_M} + \frac{1-\alpha}{\varepsilon_D} \tag{3}$$

In such medium, two possible types of waves can propagate: ordinary and extraordinary. Without loss of generality, we can assume that for all of them $k_y = 0$ due to the symmetry of the structure. The wavevector of the ordinary wave satisfies the following relation

$$k_x^2 + k_z^2 = \varepsilon_x k_0^2, \qquad k_0 = \omega/c \tag{4}$$

And the wavevector of the extraordinary wave satisfies the equation:

$$\frac{k_x^2}{\varepsilon_z} + \frac{k_z^2}{\varepsilon_x} = k_0^2 \tag{5}$$

If this metamaterial is stacked with a cladding whose dielectric constant equals to $\varepsilon_1$ (see Figure 1), a surface plasmon state will appear. If we denote the $z$-component of the wavevector in the dielectric medium as $k_{z1}$ and in the metamaterial as $k_{z2}$, then the boundary condition will give (apart from the conservation of the $k_x$ component) the following dispersion relation for an ordinary-type plasmon:

$$k_x = k_0 \sqrt{\frac{\varepsilon_1 \varepsilon_x}{\varepsilon_1 + \varepsilon_x}} \qquad (6)$$

This is similar to a conventional plasmon dispersion:

$$k_x = k_0 \sqrt{\frac{\varepsilon_1 \varepsilon_{Me}}{\varepsilon_1 + \varepsilon_{Me}}} \qquad (7)$$

For an extraordinary plasmon, the dispersion has the form:

$$k_x = k_0 \sqrt{\frac{\varepsilon_1^2 \varepsilon_z - \varepsilon_1 \varepsilon_x \varepsilon_z}{\varepsilon_1^2 - \varepsilon_x \varepsilon_z}} \qquad (8)$$

We will show now that using a metamaterial can indeed shift the LDOS peak to the low-frequency range. If a metal is modelled by the Drude theory, its dielectric constant has the form

$$\varepsilon_{Me}(\omega) = \varepsilon_0 - \frac{\omega_p^2}{\omega(\omega + i\gamma)} \qquad (9)$$

For a metal-dielectric interface (when no metamaterial is used), the frequency for the LDOS peak can be estimated when the absorption ($\gamma$) is small:

$$\omega_{peak}^{(0)} = \frac{\omega_p}{\sqrt{\varepsilon_0 + \varepsilon_1}} \qquad (10)$$

As aforementioned, since the Drude theory can be used to describe the metal in the low-frequency range, we can apply several algebraic transformations to equations 6 and 8 and get for an ordinary plasmon following:

$$\omega_{peak}^{(ord)} = \frac{\omega_p}{\sqrt{\varepsilon_0 + \varepsilon_1 + \frac{1-\alpha}{\alpha}(\varepsilon_D + \varepsilon_1)}} \qquad (11)$$

And for an extraordinary plasmon, we obtain:

$$\omega_{peak}^{(ext)} = \frac{\omega_p}{\sqrt{\varepsilon_0 + \xi \pm \sqrt{\varepsilon_1^2 + \xi^2}}}, \qquad \xi = \frac{(1-\alpha)(\varepsilon_D^2 - \varepsilon_1^2)}{2\alpha\varepsilon_D} \qquad (12)$$

It is clear that for an ordinary-wave plasmon, any value of $\alpha$ leads to a decrease in the peak frequency, while for an extraordinary one it is necessary to select "+" in "±" and satisfy $\xi > 0$ or $\varepsilon_D > \varepsilon_1$. Moreover, when $\xi = 0$ there is actually no difference in dispersion between the equation 8 and the usual equation 7, that is, using the same material for the filling and for the cladding will not lead to the appearance of a distinguishable extraordinary-wave plasmon.

Another important point is that

$$\lim_{\alpha \to 0} \omega_{peak}^{(ord,ext)} = 0 \qquad (13)$$

For the extraordinary plasmon, this is valid only when $\varepsilon_D > \varepsilon_1$. In other words, by reducing the fraction of metal, we can achieve an arbitrarily small peak frequency – without changing the materials.

In the following, we demonstrate how the proposed structure can be used to control the properties of the metamaterial, the dispersion of a surface plasmon, and the value of the Purcell factor. We will use silver as an example of metal. The value of its dielectric constant is taken from the experimental work[31]. As for dielectric, we will use several materials One of them is the organic light-emitting compound 4,4-Bis(N-carbazolyl)-1,1-biphenyl (CBP) that was previously studied for its suitability for silver-based plasmonic structures[32]. CBP is used as an example of a filling material and silica ($\varepsilon = 1.45^2$) as a cladding material.

A straightforward calculation of the dispersion of dielectric tensor components using equation 3 gives the results shown in Fig. 2. An important feature of the $\varepsilon_z$ dependence is a double peak and a change of sign at a sufficiently large value of $\alpha$.

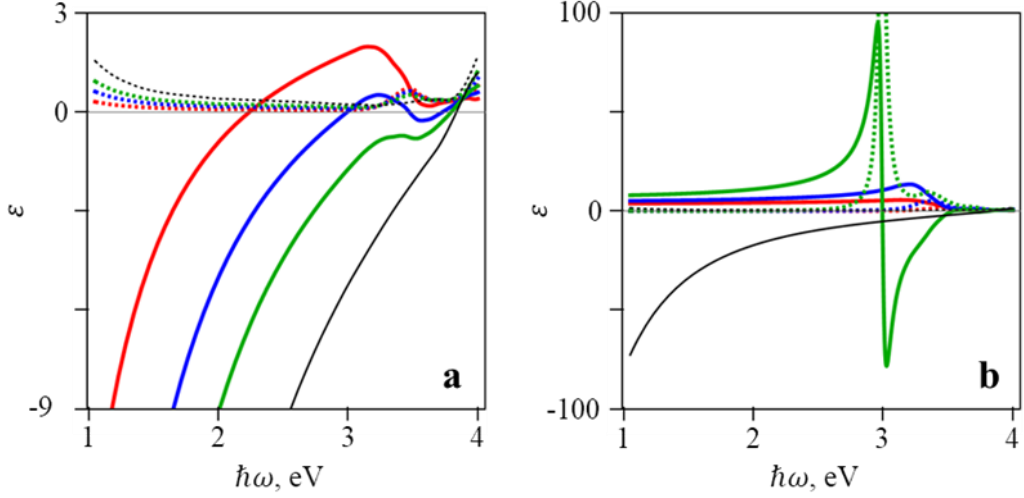

**Fig. 2:** Dependence of real (solid lines) and imaginary (dashed lines) parts of the dielectric tensor components: a) $\varepsilon_x$ and b) $\varepsilon_z$ on frequency for different values of $\alpha$ (red: $\alpha = 0.2$, blue: $\alpha = 0.4$, green: $\alpha = 0.6$) in a silver/CBP metamaterial. Black lines denote dispersion of silver dielectric constant.

The surface plasmon dispersion calculated according to equations 6 and 8 is shown in Fig. 3. For comparison, the dispersion of a simple surface plasmon at the interface between silver and both used dielectrics (CBP and silica) is also shown. Importantly, the plasmon associated with an ordinary wave is proved to be more effective in reducing the peak frequency of LDOS. It is obvious that by varying the filling factor $\alpha$ one can tune the peak frequency to any wavelength.

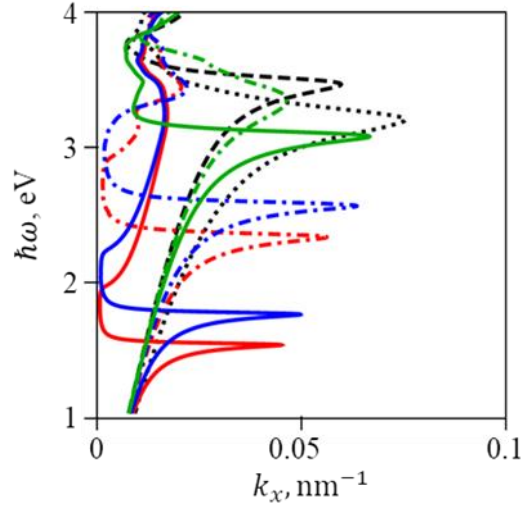

**Fig. 3:** Dispersion of the surface plasmon at the interface between silver/CBP metamaterial and silica: ordinary surface plasmon (solid lines), extraordinary surface plasmon (dash-dotted lines). Colors denote different values of $\alpha$ (red: $\alpha = 0.15$, blue: $\alpha = 0.2$, green: $\alpha = 0.7$) in a silver/CBP metamaterial. Black lines denote dispersion of a simple plasmon at the interface between silver and CBP (dotted line) and silica (dashed line).

The values of the Purcell factor have been calculated using the method described in detail previously[21]. The results are shown in Fig. 4(a), where, for comparison, the values of the Purcell factor for a simple interface plasmon are also shown, and in Fig. 4(b) where the maximum value of the Purcell factor is plotted against the filling factor $\alpha$. Clearly, lowering the LDOS peak frequency towards the low-absorption range is proven to be effective. The value of the Purcell factor can be easily increased tenfold. Importantly, a decrease in the value of the filling factor $\alpha$ increases the value of the Purcell factor for an ordinary plasmon. For very small values of $\alpha$ the maximum Purcell factor is approaching zero as there can be no surface plasmon on a dielectric-dielectric boundary (since at $\alpha = 0$ the metamaterial is nothing more than a dielectric). When $\alpha$ is increasing in the middle of its range, the value of the Purcell factor tends to decrease due to higher absorption and penetration of the electric field into the metamaterial half of the structure. Finally, at $\alpha = 1$ the maximum value of the Purcell factor is equal to the value for the conventional metal-dielectric interface plasmon as expected. Between the ordinary- and extraordinary-wave plasmons, the former is again more effective in enhancing the Purcell effect. The complex shapes of the curves in Fig. 4(b) are a consequence of the irregularities in the experimentally measured refractive index values for silver and CBP, to which the peak Purcell factor value is very sensitive.

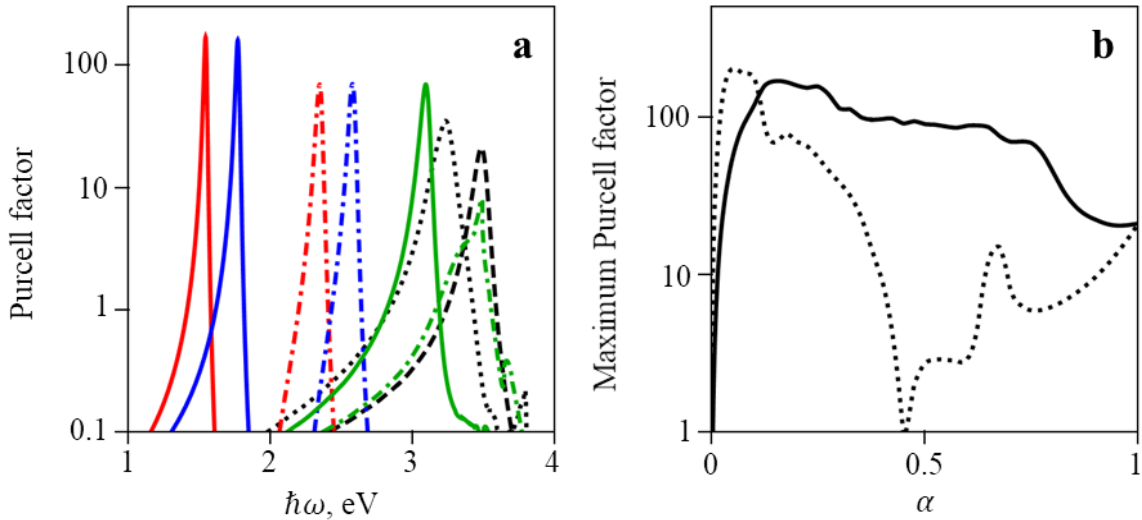

**Fig. 4:** (a) Dependence of the Purcell factor on frequency for an interface between silver/CBP metamaterial and silica. Ordinary surface plasmon (solid lines), extraordinary surface plasmon (dash-dotted lines). Colors denote different values of $\alpha$ (red: $\alpha = 0.15$, blue: $\alpha = 0.2$, green: $\alpha = 0.7$) in a silver/CBP metamaterial. Black lines denote Purcell factor of a simple plasmon for an interface between silver and CBP (dotted line) and silica (dashed line). (b) Dependence of the maximum Purcell factor on the value of $\alpha$ for the same structure. Ordinary surface plasmon (solid lines), extraordinary surface plasmon (dash-dotted lines).

To conclude, we have shown that the peak of local density states associated with the surface plasmon and the associated peak in the Purcell coefficient can be shifted towards lower energy, where the absorption of metal decreases if the bulk uniform metal is replaced by periodic metal- dielectric structures. The shift of the LDOS peak toward lower energy is accompanied by increase in the peak value of the Purcell coefficient. For the surface plasmon localized at the interface between the silver/CBP metamaterial and silica, the peak value of the Purcell coefficient can be shifted to visible/infrared energy range, where the value of the Purcell coefficient is increased by one order of magnitude.

**Acknowledgements**

The work has been supported by Russian Science Foundation grant #16-12-10503 and the Swedish Research Council (Grant 2019-05154), the Swedish Energy Agency (Grant 46563-1).

**References**

[1] R. H. Ritchie, Physical Review 106 (5), 874-881 (1957).
[2] S. A. Maier, Plasmonics: fundamentals and applications. (Springer Science & Business Media, 2007).
[3] J. B. Pendry, L. Martin-Moreno and F. J. Garcia-Vidal, Science 305 (5685), 847-848 (2004).
[4] J. A. Schuller, E. S. Barnard, W. Cai, Y. C. Jun, J. S. White and M. L. Brongersma, Nature Materials 9 (3), 193-204 (2010).
[5] A. V. Zayats, I. I. Smolyaninov and A. A. Maradudin, Physics Reports 408 (3-4), 131-314 (2005).
[6] W. L. Barnes, A. Dereux and T. W. Ebbesen, Shin Nature 424 (6950), 824-830 (2003).
[7] T. W. Ebbesen, H. J. Lezec, H. F. Ghaemi, T. Thio and P. A. Wolff, Nature 391 (6668), 667-669 (1998).
[8] A. V. Baryshev and A. M. Merzlikin, Applied Optics 53 (14), 3142 (2014).
[9] A. G. Brolo, Nature Photonics 6 (11), 709-713 (2012).
[10] J. Bellessa, C. Symonds, K. Vynck, A. Lemaitre, A. Brioude, L. Beaur, J. C. Plenet, P. Viste, D. Felbacq, E. Cambril and P.Valvin, Physical Review B 80 (3) (2009).
[11] K. Ding, M. T. Hill, Z. C. Liu, L. J. Yin, P. J. van Veldhoven and C. Z. Ning, Optics Express 21 (4), 4728 (2013).

[12] A. E. Schlather, N. Large, A. S. Urban, P. Nordlander and N. J. Halas, Nano Letters 13 (7), 3281-3286 (2013).
[13] J.-H. Song, J. Kim, H. Jang, I. Yong Kim, I. Karnadi, J., J. H. Shin and Y.-H. Lee, Nature Communications 6 (1) (2015).
[14] M. Fleischmann, P. J. Hendra and A. J. McQuillan, Chemical Physics Letters 26 (2), 163-166 (1974).
[15] B. Sharma, R. R. Frontiera, A.-I. Henry, E. Ringe and R. P. Van Duyne, Materials Today 15 (1-2), 16-25 (2012).
[16] E. M. Purcell, H. C. Torrey and R. V. Pound, Physical Review 69 (1-2), 37-38 (1946).
[17] V. P. Bykov, Soviet Journal of Experimental and Theoretical Physics 35, 269 (1972).
[18] W. Q. Huang, L. Xu and K. Y. Wuet, J. Appl. Phys. 102 (2007).
[19] I. Iorsh, A. Poddubny, A. Orlov, P. Belov and Y. S. Kivshar, Physics Letters A 376 (3), 185-187 (2012).
[20] J. B. Khurgin, Nature Nanotechnology 10 (1), 2-6 (2015).
[21] K. M. Morozov, K. A. Ivanov, D. de Sa Pereira, C. Menelaou, A. P. Monkman, G. Pozina and M. A. Kaliteevski, Scientific Reports 9 (1) (2019).
[22] A. P. Vinogradov, A. V. Dorofeenko and I. A. Nechepurenko, Metamaterials 4 (4), 181-200 (2010).
[23] J. B. Pendry, A. J. Holden, W. J. Stewart and I. Youngs, Physical Review Letters 76 (25), 4773-4776 (1996).
[24] S. Brand, R. A. Abram and M. A. Kaliteevski, Physical Review B 75 (3) (2007).
[25] C. R. Simovski, P. A. Belov, A. V. Atrashchenko and Y. S. Kivshar, Advanced Materials 24 (31), 4229-4248 (2012).
[26] A. Moroz, Physical Review Letters 83 (25), 5274-5277 (1999).
[27] M. Saboktakin, X. Ye, U. K. Chettiar, N. Engheta, C. B. Murray and C. R. Kagan, ACS Nano 7 (8), 7186-7192 (2013).
[28] X. Yu, Y. J. Lee, R. Furstenberg, J. O. White and P. V. Braun, Advanced Materials 19 (13), 1689-1692 (2007).
[29] T. Gric and O. Hess, Applied Sciences 8 (8) (2018).
[30] Y. Xiang, J. Guo, X. Dai, S. Wen and D. Tang, Optics Express 22 (3) (2014).
[31] S. Babar and J. H. Weaver, Applied Optics 54 (3), 477 (2015).
[32] K. T. Kamtekar, A. P. Monkman and M. R. Bryce, Advanced Materials 22 (5), 572-582 (2010).